\documentclass[fleqn,usenatbib]{mnras}

\usepackage{newtxtext,newtxmath}
\usepackage{ulem}

\usepackage[T1]{fontenc}

\DeclareRobustCommand{\VAN}[3]{#2}
\let\VANthebibliography\thebibliography
\def\thebibliography{\DeclareRobustCommand{\VAN}[3]{##3}\VANthebibliography}

\usepackage{graphicx}	
\usepackage{amsmath}	
\usepackage[dvipsnames]{xcolor}
\usepackage{adjustbox}
\usepackage{pdflscape}
\usepackage{float}

\title[MeerKAT MSP timing analysis]{Timing and noise analysis of five millisecond pulsars observed with MeerKAT}

\author[M.~Chisabi et al.]{M.~Chisabi,$^{1}$\thanks{E-mail: mukadichisabi@gmail.com}
S.~Andrianomena,$^{2,3}$\thanks{E-mail: andrianomena@gmail.com}
U.~Enwelum,$^{4}$ 
E. G.~Gasennelwe,$^{5}$
A.~Idris,$^{5}$\newauthor 
E. A.~Idogbe,$^{6}$
S.~Shilunga,$^{7}$
M.~Geyer,$^{8,2}$\thanks{E-mail: marisa.geyer@uct.ac.za}
D. J.~Reardon,$^{9,10}$
C. F.~Okany,$^{6,11}$\newauthor 
M.~Shamohammadi,$^{9,10}$
R.~M.~Shannon,$^{9,10}$
V.~Venkatraman~Krishnan,$^{12}$
F.~Abbate,$^{12,13}$\newauthor
M.~Kramer$^{12}$
\\
$^{1}$ Department of Physics, The Copperbelt University, Jambo Drive, Kitwe, 21696, Zambia\\
$^{2}$ SARAO, Liesbeek House, River Park Liesbeek Parkway, Settlers Way, Mowbray, Cape Town, 7705\\
$^{3}$ Department of Physics \& Astronomy, University of the Western Cape, Bellville, Cape Town 7535, South Africa\\
$^{4}$ Department of Science Laboratory Technology, University of Nigeria, Nsukka\\
$^{5}$ Department of Physics, University of Botswana, Gaborone, Botswana\\
$^{6}$ National Space Research and Development Agency, Centre for Basic Space Science, Nsukka, 410102, Nigeria\\
$^{7}$ Department of Physics, Chemistry and Material Science, University of Namibia, Private Bag 13301, Windhoek, Namibia\\
$^{8}$ High Energy Physics, Cosmology \& Astrophysics Theory (HEPCAT) Group, Department of Mathematics and Applied Mathematics,\\ \,\,  University of Cape Town,  Rondebosch 7701, South Africa\label{uct}\\
$^{9}$ Centre for Astrophysics and Supercomputing, Swinburne University of Technology, P.O. Box 218, Hawthorn, Victoria 3122, Australia\\
$^{10}$ Australian Research Council Centre of Excellence for Gravitational Wave Discovery (OzGrav)\\
$^{11}$ Department of Physics \& Astronomy, University of Nigeria, Nsukka, 410101, Nigeria\\
$^{12}$ Max-Planck-Institut f\"{u}r Radioastronomie, Auf dem H\"{u}gel 69, D-53121 Bonn, Germany\\
$^{13}$ INAF - Osservatorio Astronomico di Cagliari, Via della Scienza 5, I-09047 Selargius (CA), Italy}

\date{Accepted XXX. Received YYY; in original form ZZZ}

\pubyear{2023}

\begin{document}
\label{firstpage}
\pagerange{\pageref{firstpage}--\pageref{lastpage}}
\maketitle

\begin{abstract}
Millisecond pulsars (MSPs) in binary systems are precise laboratories for tests of gravity and the physics of dense matter. Their orbits can show relativistic effects that provide a measurement of the neutron star mass and the pulsars are included in timing array experiments that search for gravitational waves. Neutron star mass measurements are key to eventually solving the neutron star equation of state and these can be obtained by a measure of the Shapiro delay if the orbit is viewed near edge-on.
Here we report on the timing and noise analysis of five MSPs observed with the MeerKAT radio telescope: PSRs J0900$-$3144, J0921$-$5202, J1216$-$6410, J1327$-$0755 and J1543$-$5149. We searched for the Shapiro delay in all of the pulsars and obtain weak detections for PSRs~J0900$-$3144, J1216$-$6410, and J1327$-$0755. We report a higher significance  detection of the Shapiro delay for PSR~J1543$-$5149, giving a precise pulsar mass of $M_{\rm p} = 1.349^{+0.043}_{-0.061}\,$M$_\odot$ and companion white-dwarf mass $M_{\rm c} = 0.223^{+0.005}_{-0.007}$\,M$_\odot$.  This is an atypically low mass measurement for a recycled MSP. In addition to these Shapiro delays, we also obtain timing model parameters including proper motions and parallax constraints for most of the pulsars.
\end{abstract}

\begin{keywords}
pulsars: general -- stars: neutron -- astrometry -- parallaxes
\end{keywords}


\section{Introduction}

Millisecond pulsars (MSPs) are recycled neutron stars that are typically found in binary systems \citep{bhattacharya1991formation}. The timing of periodic radio pulses emitted from MSPs allows for precise measurements of the properties of the pulsars themselves and provides an opportunity for stringent tests of fundamental physics including searching for gravitational waves \citep{foster1990timing}, testing theories of gravity \citep{KramerGR}, and constraining the neutron star equation of state through mass measurements \citep{cromartie2020relativistic}. Through the detection of the Shapiro delay, a precise mass of the binary companion can be measured, and the neutron star (NS) mass inferred \citep{Shapiro64}. To date, the detection of Shapiro delay signatures have led to significant mass measurements in about $60$ binary NS systems, ranging from $M_{\rm p} \sim 1.2$\, M$_{\odot}$ \citep{ozel2016masses} to the current most massive known NS, PSR~J0740$+$6620, with $M_{\rm{p}} = 2.08(7)$ M$_{\odot}$ \citep{fonseca2021refined}. 

The measurements of the masses or radii of MSPs can strongly constrain the NS matter equation of state and, thus, the interior composition of neutron stars (e.g. \citealt{lattimer2004physics, lattimer2007neutron,Choudhury+24}). Using X-ray data from NASA’s Neutron Star Interior Composition ExploreR (NICER) instrument \citep{raaijmakers2021constraints}, onboard the International Space Station, astronomers have also determined the radii of PSRs~J0740$+$6620 and J0437$-$4715 to be $13.7^{+2.6}_{-1.5}$\,km and $11.36^{+0.95}_{-0.63}$\,km, respectively \citep{miller2021radius, Choudhury+24}. As constraints on the combination of mass and radius for more pulsars improve \citep[e.g.,][]{Reardon+24}, so too will our detailed understanding of the neutron star's equations of state and its interior composition. 

Finding pulsars with very high masses have the additional benefit that it allows for the exploration of matter on the verge of imploding to a black hole \citep{riley2021nicer}, ultimately probing the fundamental divide between neutron stars and black holes. A recent study using the MeerKAT telescope has identified a pulsar in a binary with another massive compact object. The mass of this companion lies within the range (previously considered a mass gap) between NSs and black holes; it remains uncertain whether the object is a NS or black hole \citep{barr2024}. Such discoveries motivates continued analysis of NS masses via MSP binaries.

The sensitivity of a Shapiro delay detection and the resulting precision of a pulsar's mass measurement depends on how precisely the MSP can be timed, and the magnitude of the delay signature, which is larger for binary systems viewed edge-on and with massive companions. Mass measurements are further enabled by the measurement of additional post-Keplerian parameters, such as the relativistic orbital precession (e.g. \citealt{serylak2022}). Until recently, this meant that only the brightest MSPs with favourable alignments, or highly relativistic systems, lead to NS mass estimates via Shapiro delay measurements. With the increased timing sensitivity of the MeerKAT telescope \citep{parthasarathy2021measurements, Spiewak+22} we now have the opportunity to turn to fainter or less studied MSPs in binary systems and attempt to constrain these NS masses. The MeerTime \citep{Bailes+20} relativistic binary timing programme \citep{Kramer+21} uses the MeerKAT radio telescope in the Northern Cape province of South Africa with the aim to target many such systems. 

In this work, we report on a search for Shapiro delay signatures in five MSPs observed as part of the MeerTime relativistic binary programme. The systems studied are all ideal MeerKAT targets in the Southern Sky, with declinations $\delta < -7$ deg. For all five pulsars, we describe their updated timing model parameters and noise properties.

In Section \ref{sec:data} we describe our observations from MeerKAT and summarise the data reduction pipeline. Section \ref{sec:methods} describes the noise and timing model analysis using \textsc{tempo2} and \textsc{temponest}, and Section \ref{sec:results} presents the results of this analysis. In Section \ref{sec:discuss} the results are discussed and compared with previous analyses, and the manuscript is concluded in Section \ref{sec:conclusions}.

\section{Observations}\label{sec:data}

For this work, we used MeerKAT data at L-band (856 -- 1712\,MHz) and UHF (544 -- 1088\,MHz) frequencies, obtained as part of the MeerTime Large Survey Project (LSP). 

This includes data from the relativistic binary sub-theme within MeerTime. The relativistic binary sub-theme targets pulsars which based on their known timing properties are candidate systems for new NS mass measurements via relativistic effects, or having NS masses already established are likely to provide estimates of post-Keplerian parameters through additional monitoring (for details see \citealt{Kramer+21}). These targets are typically observed by conducting orbital campaigns of these sources at both L-band and UHF. Another MeerTime sub-theme, the MeerKAT Pulsar Timing Array \cite[][MPTA]{miles2023meerkat} programme, regularly conducts shorter timing observations of MSPs at L-band with the aim of searching for nanohertz-frequency gravitational waves \citep{Spiewak+22, miles2023meerkat}. Here we utilise observations collected by both of these pulsar programmes. 

MPTA observations have typical observing durations of 256\,s (PSRs J0900$-$3144,  J1216$-$6410 and J1543$-$5149), with some weaker pulsars observed for longer, to achieve $\sim 1\,\mu$s timing precision (i.e., observations of PSR~J1327$-$0755 are typically 570\,s). Observations of PSR~J0921$-$5202 ranged from 256\,s to 2045\,s owing to a combination of MPTA observations and longer observing campaigns. For PSR~J1543$-$5149, we include observations at UHF frequencies that were obtained as part of a dedicated observing campaign following our initial detection of a potential Shapiro delay signature. These data were taken as part of the MeerTime relativistic binary timing programme. In particular, a set of 18 UHF observations were scheduled, with a 2\,hr observation coincident with the superior conjunction of the binary orbit, and two 1\,hr observations either side of the superior conjunction, to maximise our sensitivity to the Shapiro delay signature. The remaining 13 UHF observations of \mbox{$\sim$30}\,min each were distributed across orbital phase.

The MeerKAT observations for each pulsar are summarised in Table \ref{tab:obs}. MeerKAT baselines for four out of the five pulsars span just over 4 years. One of the sources in this work, PSR~J0921$-$5202, however, is no longer regularly monitored as part of the MPTA. Observing ceased because it produced poor timing precision and is therefore unlikely to contribute to the sensitivity of the MPTA to nanohertz-frequency gravitational waves. Only eight observations have been conducted for this pulsar with MeerKAT, which does not provide enough degrees of freedom to fit for the full timing model. We therefore also include published observations from the Murriyang, the 64-m radio telescope in Parkes, Australia, described by \citet{Lorimer+21}. This provided an additional 22 unique observations spanning 1.5\,yrs (MJD 56767 to 57315) and extended the baseline of our analysed dataset to just over 6\,yrs.

MeerKAT data from both receivers are recorded using the Pulsar Timing User Supplied Equipment (PTUSE) backend, which in its \textit{fold-mode} records 8 second time integrations of filterbank data folded using a known pulse period and timing ephemerides, and produces 1024 phase bins across the pulse period \citep{Bailes+20}. The data are coherently dedispersed using a fiducical dispersion measure (DM) value with a frequency resolution of 1024 channels across the observing band, or $\sim$0.8\,MHz at L-band and $\sim$0.5\,MHz for UHF. 

The data are processed using the \textsc{MeerPipe}\footnote{\url{https://github.com/OZGrav/meerpipe}} data reduction pipeline. In this pipeline, the data are summed in polarisation to total intensity, reduced to eight (L-band) or nine (UHF) frequency channels and between five and nine time sub-integrations, depending on the pulsar and observing length. Template pulse profiles were generated using the \textsc{psrchive} \citep{HotanEtAl2004} wavelet smoothing algorithm on a high signal-to-noise ratio pulse profile produced from summing all of the observations. The template profile retained the same number of channels as the observations to allow each channel to be timed against an appropriate template, accounting for profile evolution across the observing band. The times-of-arrival (ToAs) per frequency sub-band and per time sub-integration were then computed using these templates and the Fourier-domain Monte Carlo algorithm from \textsc{psrchive}. As part of the subsequent timing analysis we removed ToAs for which the associated pulse profile had a signal-to-noise ratio less than 10.

\begin{table*}
\caption{MeerKAT and Murriyang timing baselines and number of generated times-of-arrival (ToAs) at L-band and UHF frequencies as used in this work.}
\label{tab:obs}
\centering 
\begin{tabular}[width=\pagewidth]{p{0.01\textwidth} l l l l l l}
\multicolumn{2}{l}{Pulsar name} & J0900$-$3144 & J0921$-$5202 & J1216$-$6410 & J1327$-$0755 & J1543$-$5149\\
\hline
\multicolumn{2}{l}{Telescope/Backend: \textit{MeerKAT/PTUSE}}\\
& Number of ToAs at L-band & 1105 & 128 & 816 & 595 & 1010 \\
& Number of ToAs at UHF & 216  & -- & 72 & -- & 1757 \\
& Number of observations at L-band & 89  & 30 & 86 & 70 & 89 \\
& Number of observations at UHF & 3  & -- & 2 & -- & 18 \\
& Date range (MJD) & 58595 -- 60116  & 58623 -- 59086 & 58557 -- 60116 & 58855 -- 60123  & 58595 -- 60116 \\
\hline
\multicolumn{2}{l}{Telescope/Backend: \textit{Murriyang/MultiBeam}}\\
& Number of ToAs at 1382\,MHz & -- & 22 & -- & -- & -- \\
& Number of observations at 1382\,MHz & -- & 22 &-- &-- & --\\
& Date range (MJD) & -- & 56767 - 57315 & -- & --  & -- \\
\hline
\end{tabular}
\end{table*}

\section{Methods}\label{sec:methods}

\subsection{Timing and noise models}

We use standard timing and noise modelling techniques to establish robust timing model parameters.
The ToAs   are compared to timing models that describe the spin evolution and astrometric properties of the pulsar, the orbit if it has a binary companion and the line-of-sight propagation through the ionised interstellar medium (ISM). Relativistic corrections to the pulse propagation through the surrounding curved space-time, in the case of a binary system, can also be modelled, enabling measurements of the pulsar and companion masses. 
Arrival times were corrected from an observatory time standard to the BIPM(2021) realisation of International Atomic Time\footnote{https://webtai.bipm.org/ftp/pub/tai/ttbipm/TTBIPM.2021}.
Thereafter, arrival times were transformed to the solar system barycentre using the DE440 Solar System ephemeris as obtained from the Jet Propulsion Laboratory (JPL; \citealt{Park2021}). For all pulsars, except PSR J1327$-$0755 as discussed in Sec. \ref{sec:psrj13}, we assume the default line-of-sight solar wind electron density of $n_e = 4\,{\rm cm}^{-3}$.

The parameters describing the timing model undergo a least-squares fit of the linearised model to the measured ToAs using \textsc{Tempo2} \cite[][]{Edwards+06}. The difference between the observed ToAs and those predicted by the pulsar timing model are the \textit{timing residuals}. The timing residuals contain the latent measurement errors, pulse jitter, timing noise (e.g. \citealt{shannon2010assessing}), and any unmodeled physical or instrumental processes. These noise processes need to be modeled to  enable a robust generalised least-squares fit of the timing model parameters, which returns reliable parameter uncertainty estimates. 

We use the ``ELL1" binary model appropriate for the near-circular orbits of the pulsars in our sample \citep{Lange2001}. This model ignores orbital perturbations that depend on terms with eccentricity ($e$) of the order $e^2$ or higher. We confirm that these additional delays, $\delta t_{\rm ell1} \sim 0.5 x e^2$, for $x$ the projected semi-major orbital axis, are much smaller than the weighted root-mean-square timing residual for each pulsar we studied. We also include, for all pulsars an ``FD" parameter that models residual frequency-dependent delays caused, for example, by profile template errors \citep{Zhu2015}.

\subsection{Noise modelling}
\label{sec:noisemodelling}

We use \textsc{TempoNest} \citep{temponest} to characterise the pulsar timing model while simultaneously assessing for the presence, and modelling the effect of, additional stochastic noise processes.
\textsc{TempoNest} can conduct a simultaneous Bayesian analysis of the linear or non-linear deterministic pulsar timing model and additional stochastic parameters. It uses the Bayesian inference nested sampler \textsc{MultiNest} \footnote{\href{http://ccpforge.cse.rl.ac.uk/gf/project/multinest/}\href{http://ccpforge.cse.rl.ac.uk/gf/project/multinest/}} to efficiently explore the noise model parameter space \citep{feroz2009multinest}. We also use \textsc{TempoNest} to sample over the timing model parameters of most importance for this work, which are those describing the curved spacetime, and the pulsar timing parallax, while analytically marginalising over other parameters using the design matrix derived from a   \textsc{tempo2} fit. 

The noise model includes known stochastic processes, such as white noise processes to account for pulse jitter, dispersion measure variations, and achromatic red noise. Within the \textsc{TempoNest} framework the stochastic white noise processes are defined by the parameters, EFAC, EQUAD and ECORR. The EFAC, $F$, parameters apply a scaling factor per observing backend to all ToA uncertainties to account for errors arising from the template matching procedure by which the ToA uncertainties were obtained. The EQUAD, $Q$, parameter is a secondary white noise parameter, which accounts for an independent source of noise, added in quadrature to the scaled ToA uncertainties. The modified uncertainty, $\sigma_{\rm mod}$, depends on the original uncertainty, $\sigma_{\rm TOA}$, as $\sigma_{\rm mod} = \sqrt{(F \sigma_{\rm TOA})^2 + Q^2}$ \citep{temponest}. Finally the ECORR parameter describes a process that is also added in quadrature but that is fully correlated in frequency for multi-band ToAs, but independent in time \citep{ng9yrdata}. It therefore captures, for example, frequency dependent profile errors weighted by flux density variations from scintillation, pulse jitter, or propagation effects within a given observing epoch. 

Slow varying changes in (observing frequency-dependent) DM values are modeled by a long-term time correlated noise process, with its power spectral density (PSD) described by a power law, and a Fourier basis for which the number of Fourier frequency components are modeled to a maximum frequency of 1/30\, days$^{-1}$, or roughly twice our observing cadence. An additional red noise process, with a similar PSD, is also modeled to account for either spin irregularities in the pulsars, or for contributions from the putative stochastic gravitational-wave background \citep{ng23, epta23, ppta23, cpta23}. These power law PSDs are described by the parameters DMAmp, DMSlope  and RNAmp and RNSlope in \textsc{TempoNest} respectively. 

In our analysis we include all noise model parameters for all pulsars. We quantify the data-driven Bayesian support for each of these processes via the Savage-Dickey (SD) density ratio, which gives an estimate of the Bayes factor (BF, or $\mathcal{B}$) for each process \cite[see][]{Arzoumanian+18}. This Bayes factor is calculated by comparing the posterior support for a noise process in the weak signal regime  to the prior density. 

\subsection{Searching for Shapiro delays}

The Shapiro delay \citep{Shapiro64} is the retardation of the pulse arrival times from a pulsar  when pulses propagate through the compressed space-time associated with a binary companion. The Shapiro delay varies over binary orbital phase, with the largest delay observed during superior conjunction when the pulsar is directly behind its companion. The Shapiro delay can be characterised by two parameters, the range,

\begin{equation}
r = T_\odot\, M{_{\rm c}} \\
\end{equation}

and shape
\begin{equation}
s = \sin i ,
\end{equation}
\noindent where $M_{\rm{c}}$ is the mass of the companion, $i$ the inclination of the orbit on the sky, and $T_\odot = G M_\odot /c^3 = 4.925490947 \, \mu$s is the mass of the Sun in units of time.

These Shapiro delay quantities can alternatively be expressed in terms of pair of orthometric parameters \citep{FreireAndWex2010}, namely the harmonic amplitude ($h_3$)  

\begin{equation}\label{eq:h3}
h_3 = T_{\odot} \,M_{\rm{c}} \,\varsigma^3,
\end{equation}
and ratio ($\varsigma$)
\begin{equation}\label{eq:stig}
\varsigma = \frac{\sin i}{1 + |{\cos i}|}.
\end{equation}
\noindent  It is particularly advantageous to search for Shapiro delay signatures in approximately circular binaries using this less covariant parameterization.
For each pulsar analysed here we have low values of eccentricity and therefore search for Shapiro delays by sampling the orthometric parameters in \textsc{temponest}, simultaneously with the noise model. Once obtained from Bayesian inference, the Shapiro delay parameters ($h_{3}$, $\varsigma$) are used to estimate the mass of the companion ($M_{\rm c}$) and the orbital inclination ($i$) from equations \ref{eq:h3} and \ref{eq:stig}, while the pulsar mass is derived using the mass function,
\begin{equation}
    \frac{(M_{\rm c}\sin i)^3}{(M_{\rm p}+M_{\rm c})^2} = \frac{4\pi^2 \, x_{\rm p}^3}{T_{\odot}\, P_{\rm b}^2},
\end{equation}

with $P_{\rm b}$ and $x_{\rm p}$ the orbital period and the projected semi-major axis respectively, and $T_{\rm \odot} = \frac{G M_{\odot}}{c^3} =  4.925490947\, \upmu$s.  

\section{Results}\label{sec:results}  

\begin{figure*}
\includegraphics[width=2\columnwidth]{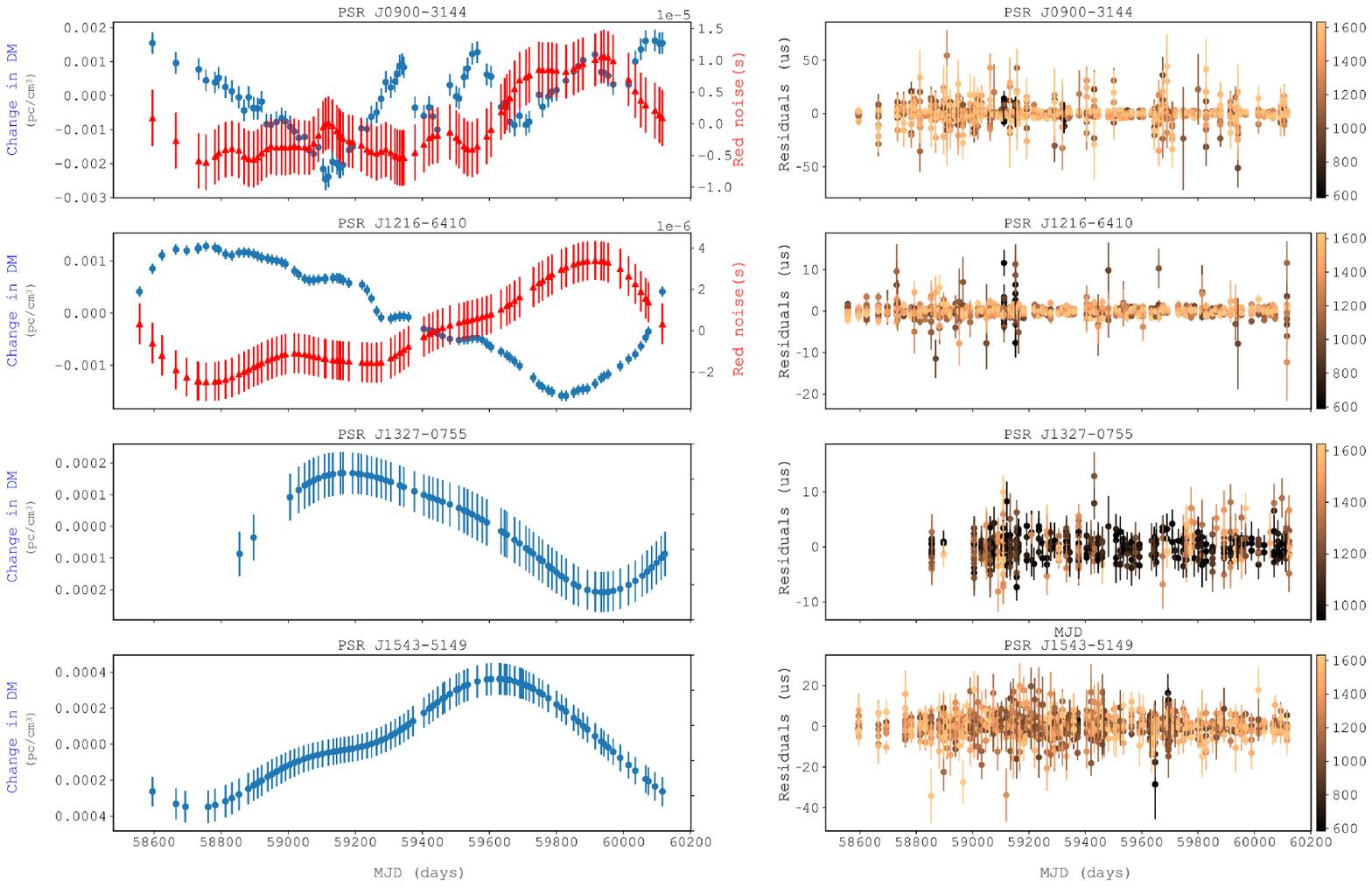}
\caption{
{\em Left column:} Variations in dispersion measure values (in pc cm$^{-3}$, in blue) and timing residuals due to red noise (in $s$, in red) across observing days, for the pulsars in our sample for which we found evidence for DM and/or Red noise.  The plotted noise model for PSR~J1543-5149 does not include red noise since it is poorly constrained with only modest support (see Table \ref{tab:timing}). {\em Right column:}  Noise subtracted residuals (in $\upmu$s) for our data sets. Each marker is coloured according to the frequency bin (which is denoted by the colour bar) it belongs to. The observations of PSR J1543-5149 (bottom row) include both L and UHF-bands.}
\label{fig:pulsarwithdmrn}
\end{figure*}

\begin{figure*}
\includegraphics[width=1.5\columnwidth]
{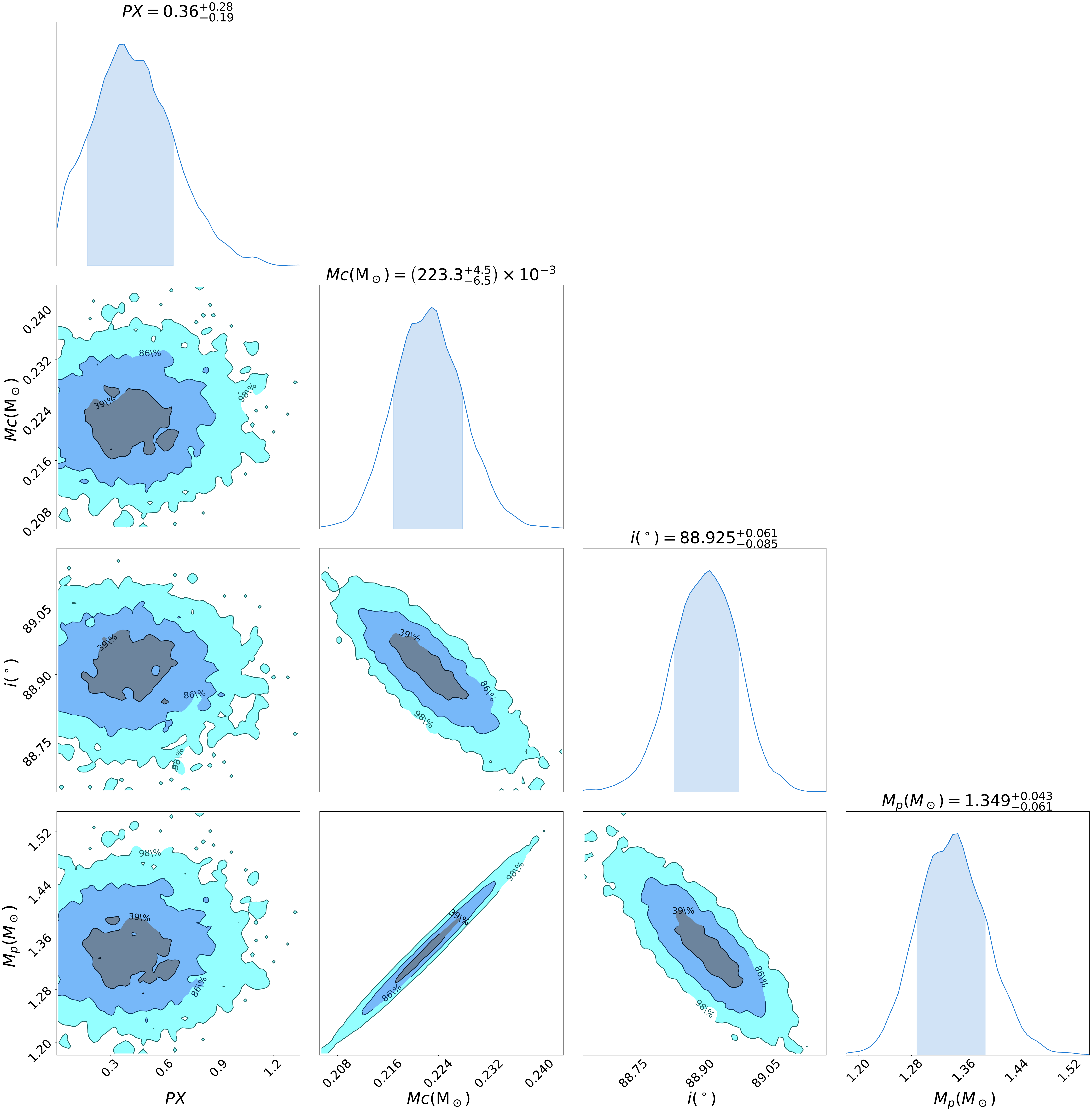}
\caption{Posterior distributions for a subset of timing parameters of PSR J1543$-$5149. Here we show parallax (PX) as well as mass of the companion ($M_{\rm c}$) and pulsar ($M_{\rm p}$) and the inclination angle of the binary. Note that the sense of the inclination angle is not established, therefore $180^\circ - i$ is an equally likely estimate to the value of $i$ presented here.}
\label{fig:1}
\end{figure*}

We discuss the characteristics and analysis for each MSP independently, reporting updated timing parameters  and properties of the preferred noise models for each. For all pulsars (except PSR J0921$-$5202 for which we have very little data) we evaluated evidence for white noise, red noise and DM noise processes as discussed in Sec \ref{sec:noisemodelling}. We include white noise parameters for all studied systems, and DM and Red-noise parameters where we find sufficient evidence for them as assessed using Savage-Dickey density ratios. We also searched for parallax and Shapiro delay parameters, $h_3$ and $\varsigma$ for all pulsars.
A summary of the obtained timing parameters is presented in Table~\ref{tab:timing}.

\begin{table*}
\caption{Timing parameters for our pulsar sample. We assume the DE440 solar system ephemeris and the BIPM2021 time standard for all pulsars. Symmetric errors are presented in parentheses and apply to the last quoted decimal place, whereas asymmetrical errors as obtained from Bayesian posterior analysis or Monte Carlo simulation are explicitly stated. All uncertainties represent $1\sigma$, $68$\% confidence interval. The transverse velocities$^\dagger$ are derived using our measured pulsar distance, where available, or with a distance estimate (with 20\% error bar) from a Galactic electron density model \citep{YMW17} otherwise.}
\centering
\begin{adjustbox}{width=1\textwidth}
\begin{tabular}[width=0.9\pagewidth] {l l l l l l}
\hline
Pulsar & J0900$-$3144 & J0921$-$5202 & J1216$-$6410 & J1327$-$0755 & J1543$-$5149\\
\hline
\multicolumn{5}{l}{Observation and reduction parameters}\\
\hline
Solar System ephemeris\dotfill & DE440 & DE440 & DE440 & DE440 & DE440 \\[0.2ex]
Time standard\dotfill & TT(BIPM2021) & TT(BIPM2021) & TT(BIPM2021) & TT(BIPM2021) & TT(BIPM2021) \\[0.2ex]
Reference epoch (MJD)\dotfill &59157& 57926 & 59143 & 59292 & 59163\\[0.2ex]
Solar wind electron density, $n_{e}$ (cm$^{-3}$)\dotfill &4 &4 &4 & 6.7(6)&4\\[0.2ex]
\hline
\multicolumn{5}{l}{Spin and astrometric parameters}\\
\hline
Period and position epoch\dotfill & 59157 & 57926 & 59143 & 59292 & 59163\\[0.2ex]
Right ascension, $\alpha$ (J2000, h:m:s)\dotfill & 09:00:43.95221(5)& 09:21:59.8780(6)  & 12:16:07.321166(14)  & 13:27:57.58064(9) & 15:43:44.145260(10)\\[0.2ex]
Declination, $\delta$ (J2000, d:m:s)\dotfill & $-$31:44:30.8723(7)  & $-$52:02:38.448(5) & $-$64:10:09.16646(10) & $-$07:55:29.912(4) & $-$51:49:54.71366(16)\\[0.2ex]
Proper motion in $\alpha$, $\mu_{\alpha}$ (mas\,yr$^{-1}$)\dotfill & $-$1.2(5)  & $-$5.2(18) & $-$7.87(8) & $-$2.5(1.2) & $-$3.96(8)\\[0.2ex]
Proper motion in $\delta$, $\mu_{\delta}$ (mas\,yr$^{-1}$)\dotfill & 1.8(5) & 8.8(18)  & 2.57(8)  & 8(3) & $-$2.36(13)\\[0.2ex]
Parallax, $\varpi$ (mas)\dotfill & $0.47^{+1.04}_{-0.41}$ & --  & 1.14$^{+0.46}_{-0.44}$  & 0.66(22) & $0.36^{+0.28}_{-0.19}$\\[0.2ex]
Spin frequency, $\nu$ (Hz)\dotfill & 90.011841777132(10) & 103.30777266468(4)  & 282.535701428469(7)  & 373.423698381884(5) & 486.154230881221(9)\\[0.2ex]
Spin-down rate, $\dot{\nu}$ ($10^{-16}$\,Hz\,s$^{-1}$)\dotfill &  $-$3.954(6)  & $-$1.928(7) & $-$1.276(4) & $-$2.519(3) & $-$3.8205(5)\\[0.2ex]
Dispersion measure, DM (cm$^{-3}$\,pc)\dotfill & 75.6866(3) & 122.711(3)  & 47.39301(15) & 27.9084(5) & 50.98325(12)\\[0.2ex]
\hline
\multicolumn{5}{l}{Binary model parameters}\\
\hline
Model name\dotfill & ELL1H & ELL1  & ELL1H  & ELL1H & ELL1H\\[0.2ex]
Orbital period, $P_{\text{b}}$ (days)\dotfill & 18.7376360567(17) & 38.22367666(11) & 4.03672725263(11) & 8.4390861450(5) & 8.0607731359(5)\\[0.2ex]
Projected semi-major axis, $x_{\rm p}$ (lt-s)\dotfill & 17.24880847(18)  & 19.053379(6)  & 2.93709185(6) & 6.64577327(13) & 6.48028675(8)\\[0.2ex]
Epoch of ascending node $T_{\rm asc}$\dotfill & 52678.6302890(7)  & 56769.836938(4) &53055.36126059(16)  & 54717.3832595(3) & 54929.0678258(3)\\[0.2ex]
First Laplace-Lagrange parameter, $\epsilon_1$\dotfill & 9.86(3)$\times 10^{-6}$ & $-$8.9(8)$\times 10^{-6}$ & 1.6(4)$\times 10^{-7}$  & $-$7.6(36)$\times 10^{-8}$ & 2.047(3)$\times 10^{-5}$ \\[0.2ex]
Second Laplace-Lagrange parameter, $\epsilon_2$\dotfill & 3.50(3)$\times 10^{-6}$  & 1.11(8)$\times 10^{-5}$ & $-$5.61(4)$\times 10^{-6}$  & $-$4.6(36)$\times 10^{-8}$ & 6.010(20)$\times 10^{-6}$\\[0.2ex]
\hline
\multicolumn{5}{l}{Noise parameters}\\
\hline
DM noise amplitude ($\log_{10}$ A$_{\rm DM}$) \dotfill  & $-$10.857$^{+0.069}_{-0.067}$  & --  & $-$11.247$^{+0.085}_{-0.076}$  & $-$11.78$^{+0.36}_{-0.60}$ & $-$11.54$^{+ 0.40}_{- 0.20}$\\[0.2ex]
DM noise spectral index\dotfill & 1.65$^{+0.20}_{-0.27}$& --  & 3.10$^{+0.40}_{-0.39}$ & 3.1$^{+1.9}_{-1.2}$ & 2.42$^{+ 0.59}_{- 1.95}$\\[0.2ex]
Red noise amplitude ($\log_{10}$ A$_{\rm{Red}}$) \dotfill & $-$12.143$^{+0.062}_{-0.109}$  & --  & $-$12.87$^{+0.11}_{-0.17}$& --  & $-$13.34$^{+0.58}_{-0.89}$\\[0.2ex]
Red noise spectral index\dotfill & 2.58$^{+0.42}_{-0.49}$  & --  & 3.87(86) & -- & 4.01$^{+ 1.20}_{- 1.69}$\\[0.2ex]
Weighted timing residual, w$_{\rm{rms}}$ ($\mu$s)\dotfill &7.768 &23.7 &1.805& 1.507&2.304\\[0.2ex]
Weighted residuals noise-subtracted w$_{\rm{rms}}$ ($\mu$s)\dotfill&1.960& -- &0.774&1.421 &1.685\\[0.2ex]
\hline
\multicolumn{5}{l}{Shapiro delay, mass and inclination measurements}\\
\hline
Orthometric amplitude, $h_3$ ($10^{-7}$ s)\dotfill &4.6$^{+1.6}_{-1.5}$ & -- & $0.71^{+0.36}_{-0.39}$& $0.97^{+0.91}_{-0.65}$& $10.4^{+0.19}_{-0.26}$\\
Orthometric ratio, $\varsigma$\dotfill  & --&  -- & -- & --& $0.9814^{+0.0011}_{-0.0015}$\\
Orbital inclination, $i$ ($\deg$)\dotfill & --  &-- & --  & -- & $88.925^{+0.061}_{-0.085}$\\[0.2ex]
Companion mass, $M_{\text{c}}$ (M$_{\odot}$)\dotfill & -- & -- & -- &  -- &$0.2233^{+0.0045}_{-0.0065}$\\[0.2ex]
Pulsar mass, $M_{\rm p}$ (M$_{\odot}$)\dotfill & -- & -- & -- & -- & $1.349^{+0.043}_{-0.061}$\\[0.2ex]
\hline
\multicolumn{5}{l}{Derived parameters}\\
\hline

Total proper motion, $\mu_{\text{T}}$ (mas\,yr$^{-1}$)\dotfill&2.2(5) & 10.22(18) & 8.28(8) &8(3)  &  4.61(14)\\
Spin period, $P_{0}$ (ms)\dotfill & 11.109649355648(1) & 
9.679813766248(4) & 3.53937571409245(90)  & 2.67792323929411(4) & 2.05696039750053(4) \\
Spin period derivative, $\dot{P}$ ($10^{-20}$\,s\,s$^{-1}$)\dotfill& 4.880(8) & 1.807(7) & 0.1598(5) &0.1806(3)  & 0.16165(2)\\
Distance estimate (1/$\varpi$, kpc) \dotfill & -- & --& $0.88^{+0.55}_{-0.23}$  & $1.5^{+0.8}_{-0.4}$  & $2.8^{+3.1}_{-1.6}$\\
Dispersion distance (kpc) \dotfill & 0.38 & 0.36 & 1.10 & $<$25 & 1.15 \\
Transverse velocity$^{\dagger}$ ($v_\perp$, km/s) \dotfill & 4 & 17& $34^{+22}_{-9}$  & 57$^{+30}_{-15}$  & $61^{+68}_{-35}$\\
[0.2ex]
\hline
\hline
\label{tab:timing}
\end{tabular}
\end{adjustbox}
\end{table*}

\subsection{PSR~J0900$-$3144}
\label{sec:J0900}

PSR~J0900$-$3144 was discovered in the Parkes high-latitude pulsar survey \citep{Burgay+06}. It continues to be regularly monitored by the European Pulsar Timing Array (EPTA) programme. A previous timing analysis of this pulsar was conducted using 7~years of observations from the EPTA \citep{Desvignes+16}. The pulsar is also observed by the Parkes Pulsar Timing Array and a noise analysis has been conducted \citep{Reardon+23}.

Using 2 years and 9 months of observations from MeerKAT, we
detect significant dispersion measure variations, DM$(t)$ and also find strong evidence for red noise as shown in the top left panel in Figure~\ref{fig:pulsarwithdmrn}.  During an initial inspection of the timing residuals using a timing model that was fitted to the first $\sim$2\,yrs of data without accounting for red noise, we identified structure in the residuals that resembles a weak pulsar glitch occurring near the beginning of 2023. Although rare, glitches have been observed in millisecond pulsars \citep{McKee2016, Cognard+04}. Using \textsc{TempoNest}, we searched for glitch parameters in this pulsar, but found no significant evidence compared with a model of just red noise. We find an insignificant Bayes Factor ($\ln{\mathcal{B}} = 1.1$, inferred by comparing the evidences computed with nested sampling) for the model with both red noise and a glitch relative to a model of only red noise. The inferred glitch amplitude, corresponding to a change in spin frequency, is $(1.5 \pm 0.7)\times 10^{-10}$\,Hz. Using our data we therefore conclude that the appearance of glitch-like residuals may be due to stochasticity in the red noise and that additional data is required to confirm or refute a glitch.

We find only weak evidence for correlated noise (ECORR) in the L-band observing system, with a Savage-Dickey density ratio of SD = 1.3, but no evidence for correlated noise at UHF (SD = 0.69). Our best-fitting timing residuals, after subtracting the characterised noise processes, are plotted in the top right panel of Figure~\ref{fig:pulsarwithdmrn}.

When we searched for a Shapiro delay we were able to constrain $h_3$, as reported in Table \ref{tab:timing}, but  unable to constrain $\varsigma$. We are therefore neither able  to constrain the pulsar mass or orbital inclination for PSR~J0900$-$3144. We report a marginal detection of timing parallax of $\varpi = 0.47^{+1.04}_{-0.41}$ mas.

We measure a proper motion of $\mu_\alpha = -1.2(5)$ mas yr$^{-1}$ in right ascension and $\mu_\delta = 1.8(5)$ mas yr$^{-1}$ in declination. In \citet{Desvignes+16} the authors present proper motions of $\mu_\alpha = -1.01(5)$ mas yr$^{-1}$ and $\mu_\delta = 2.02(7)$ mas yr$^{-1}$. These measurements are consistent with ours, however the earlier measurements of \citet{Desvignes+16} are much more precise given their longer data span of almost 7 years.
Our astrometry  is also consistent with \citet[][]{shamohammadiastro} which presented astrometry from  a nearly identical MeerKAT data set for this pulsar.

\subsection{PSR~J0921$-$5202}

This target was first discovered in archival data from the Parkes Multibeam Pulsar Survey \citep{mickaliger2012discovery}. A timing analysis spanning approximately 3~years followed using Murriyang \citep{lorimer2021timing}. From this analysis the authors placed an upper limit on the total proper motion for PSR~J0921$-$5202 of $< 216$ mas yr$^{-1}$. 
In our analysis, we use the available MeerKAT L-band data (taken between 2019 May 20 and 2020 Aug 24), to which we add $\sim1.5$yr historic Murriyang data.  We account for the constant timing offsets between these data sets by including an additional jump parameter in \textsc{tempo2}. We find no evidence for DM, red or correlated white noise in our data set, and therefore only include the white noise parameters EFAC and EQUAD in our analysis. 
We also do not find any evidence for parallax. We measure proper motions for this pulsar for the first time, finding $\mu_\alpha \cos \delta = -5.2(1.8)$ mas yr$^{-1}$ in right ascension and \mbox{$\mu_\delta = 8.8(1.7)$ mas yr$^{-1}$} in declination, which provides a total proper motion of  $\mu_{\rm T} = 10.22(18)$ mas yr$^{-1}$.

\subsection{PSR~J1216$-$6410}

This pulsar was first discovered during the Parkes Multibeam Pulsar Survey (PMPS,  \citealt{lorimer2006parkes}). It has the tightest orbit of our analysed sample, with an orbital period of 4.04 days. As part of the PMPS, PSR~J1216$-$6410  was observed for 2.1 years. With 3.2 years of total observing time using the MeerKAT L-band, we measure a proper motion of  $\mu_{\alpha} = -7.87(8)$ mas yr$^{-1}$ in right ascension and \mbox{$\mu_\delta = 2.57(8)$ mas yr$^{-1}$} in declination, which provides a total proper motion of \mbox{$\mu_{\rm T} = 8.28(8)$ mas yr$^{-1}$}. We furthermore obtain a parallax measurement of  $\varpi = 1.14^{+0.46}_{-0.44}$ mas, which translates to a distance estimate of $0.88^{+0.55}_{-0.23}$\,kpc, when using $1/\varpi$ as the conversion. 
These are consistent with the astrometry presented in \citet[][]{shamohammadiastro}.

We find clear evidence for both a DM noise process and a red noise process in our dataset, but no evidence for jitter noise (all ECORR parameters have SD ratios below 0.7). We note that the maximum likelihood time-domain realisations of the DM and red noise processes in Figure \ref{fig:pulsarwithdmrn} are anti-correlated. This can arise because of other unmodelled chromatic red processes, such as scattering variations \citep{Lentati+16}. However, since our noise model whitens the timing residuals sufficiently to provide accurate timing model measurements, we defer a detailed examination of the noise processes in this pulsar to future work.

We are able to constrain the amplitude of $h_3$, as indicated in Table \ref{tab:timing}, but not $\varsigma$. 

\subsection{PSR~J1327$-$0755} \label{sec:psrj13}

PSR~J1327$-$0755 was discovered as part of a drift scan undertaken by the Green Bank Telescope (GBT) at 350\,MHz while immobilized during track refurbishments \citep{Boyles+13}. Using $70$ observations, and over 3.5 years of observation at MeerKAT L-band, we detect a parallax of $\varpi = 0.66 \pm 0.22$ mas. This measurement deviates from the weak detection ($\varpi=0.13^{+0.17}_{-0.09}$\,mas) presented in \citet[][]{shamohammadiastro}, but is more consistent with the DM distance estimate from the NE2001 Galactic electron density model which predicts, $D=1.73$\,kpc \citep{ne2001}, while \citep{YMW17} places the pulsar much more distant $D=25$\,kpc. The discrepancy between our measurement and \citet[][]{shamohammadiastro}, likely stems from our longer time span (3.5 yr compared to their 2.4 yr), allowing us to obtain a more accurate  $\varpi$ estimate.

We measure a proper motion of $\mu_\alpha = -2.5(1.2)$ mas yr$^{-1}$ in right ascension and $\mu_\delta = 7.5(2.9)$ mas yr$^{-1}$ in declination, consistent with the measurements reported in \citet[][]{shamohammadiastro}. We searched for a Shapiro delay, and constrained only $h_3$ very weakly.

We find moderate evidence for dispersion measure variations as captured in the DM noise parameters in Table \ref{tab:timing}, with a SD density ratio of 6.7 and find no evidence for a red noise process (SD density of 0.7).

This pulsar has a low ecliptic latitude (1.2 degrees) and is therefore more sensitive to variations in the solar wind than the other pulsars in our sample. For this reason, we fit for the mean solar wind electron density at a distance of 1\,AU, and find $n_e = 6.7 \pm 0.6 \,{\rm cm}^{-3}$. This mean density is used to compute the dispersive time delays due to the solar wind at each observing epoch, assuming a spherically-symmetric $n_e$ distribution \citep{Edwards+06}.

\subsection{PSR~J1543$-$5149}

PSR J1543$-$5149 was discovered as part of the High Time Resolution Universe (HTRU) survey for pulsars and fast transients carried out from 2008 with Murriyang \citep{Keith2011}.

The pulsar, with its spin period of 2.06 ms, is the fastest spinning pulsar in our data set. It has an orbital period of 8.06 days and was first detected with a S/N of 16 in a Murriyang observation offset 0.08$^\circ$ from the nominal pulsar position. It is in a binary system with a white dwarf companion that has a minimum companion mass of 0.22\,M$_{\odot}$, making it a low-mass pulsar binary \citep{Keith2011}.

 MeerKAT observed PSR~J1543$-$5149 for 6.4 hrs in the L-band, as well as 1.3 hrs in UHF. We detected significant DM variations as shown in the bottom panel of Figure~\ref{fig:pulsarwithdmrn}, but no correlated noise (ECORR) in either the L-band or UHF band observing systems. We find modest evidence for red noise, with an estimated SD ratio of 2.3. We find no significant parallax for this pulsar.
 
We detect a Shapiro delay signature for PSR J1543$-$5149, with measured values $h_3 = 1.04^{+0.19}_{-0.25}\mu$s and $\varsigma = 0.9814^{+0.0011}_{-0.0015}$. This allows us to constrain the mass of the pulsar to $M_{\rm p} =  1.349^{+0.043}_{-0.061} M_{\odot}$, the companion mass to $M_{\rm c} = 0.2233^{+0.0011}_{-0.0015} M_{\odot}$, and the orbital inclination angle to $i = 88.925^{+0.061}_{-0.085}$ degrees. It is because of this near edge-on orbital orientation that we are able to clearly distinguish the Shapiro-delay signature for this pulsar. The posterior probability distributions of $h_3$ and $\varsigma$, as well as $M_{\rm c}$, $M_{\rm p}$, and $i$ that are derived from $h_3$ and $\varsigma$, are plotted in Figure 2. 

We measure a proper motion of $\mu_\alpha = 3.85 \pm 0.12 $ mas yr$^{-1}$ in right ascension and $\mu_\delta = -2.86 \pm 0.18$mas yr$^{-1}$ in declination. Our measurement of $\mu_\alpha$ is in modest ($\approx 1\sigma$) disagreement with the value presented in \cite{shamohammadiastro}, likely because of different noise models.
We also find evidence for a parallax measurement, obtaining $\varpi = 0.36^{+0.28}_{-0.19}$ mas, as presented in Figure 2, and consistent with \citet[][]{shamohammadiastro}. This mildly constrained value leads to a distance estimate of PSR J1543$-$5149 of $2.8^{+3.1}_{-1.6}$\,kpc, when using $1/\varpi$. All additional measured timing model parameters including proper motion are shown in Table \ref{tab:timing}.

\section{Discussion}\label{sec:discuss}

\subsection{The low mass of PSR~J1543$-$5149}

Our inferred mass of $M_{\rm p} = 1.349^{+0.043}_{-0.061}$ is unusually low for a recycled MSP. There are only five known MSPs with mass measurements constrained to be below $M_{\rm p} = 1.4\,$M$_\odot$ with at  least 1$\sigma$ significance. The other MSPs (defined as $\nu > 100\,$Hz) with such masses are PSRs~J0514$-$4002A, J1807$-$2500B, J1918$-$0642 and J2234+0611 \citep{Ridolfi2019,Lynch2012,Arzoumanian2018, Stovall2019}. These pulsars are mostly in eccentric orbits (PSR~J2234+0611) or in orbits formed through exchange encounters in a globular cluster (PSRs~J0514$-$4002A and J1807$-$2500B) and as such their masses can depend on a complex evolutionary history. PSRs J1918$-$0642 and ~J1543$-$5149 stand out among these low-mass MSPs as they are in standard highly circular orbits with a Helium white-dwarf companion and are likely to have formed through the most common MSP evolution channel. PSR J1918$-$0642, while centered a lower mass value of $M_{\rm p} = 1.29^{+0.10}_{-0.09}$ has a less precise measurement than our measurement for PSR J1543$-$5149. We estimate the accreted mass during this MSP evolution to be $\Delta M \sim 0.1$ M$_\odot$ from Equation 14 of  \citet{Tauris+12}, suggesting a birth neutron star mass of $M_{\rm p} \sim 1.25$ M$_\odot$.

\subsection{Pulsar and companions mass estimates}

For each pulsar with a weak detection of Shapiro delay (significant $h_3$ only), PSRs J0900$-$3144, J1216$-$6410, and J1327$-$0755, we use the numerical model prescribed in \cite{shamohammadi2022searches} to predict how the uncertainty of the pulsar mass may improve with further observations. 

For each pulsar, we use our measured binary mass function and assume that the inclination angle is $i=60^\circ$, which is the median for a uniform distribution. The pulsar masses, which heavily depend on the inclination angle, are poorly constrained in this scenario. Continued monitoring of these pulsars is therefore unlikely to constrain the fractional uncertainty to less than  one third.

Using the $\chi^2$ analysis in \citet[][Section 4.1]{shamohammadi2022searches}, we set $M_{\rm p}=1.2$ M$_\odot$ for PSRs~J0900$-$3144, J1216$-$6410, and J1327$-$0755 to find the $16$\%, $50$\%, and $84$\% percentiles of $\cos i$, which are used to constrain the companion mass and the orbital inclination angle for each of these systems. We repeated the analysis for $M_{\rm p}$ equal to $2.0$ M$_\odot$. The results are shown in Table \ref{tab:mass_inclination}. 

Based on the companion mass ranges obtained, we note the companion mass for PSR J0900$-$3144, $M_{\rm c}> 0.38 M_{\odot}$ for all sampled values of pulsar mass, and at the lowest end of the pulsar mass range, i.e. at $M_{\rm p} = 1.2 M_{\odot}$, the obtained mean value $\overline{M_{\rm c}} = 0.60 M_{\odot}$. This mean value is larger than the typical mass associated with a He white-dwarf (WD) companion. We conclude that PSR J0900$-$3144 likely instead has a CO-WD companion \citep{Tauris1999}. In the case of PSR J1216$-$6410 we find that for the lowest pulsar mass, the associated $\overline{M_{\rm c}} = 0.28 M_{\odot}$, typical of a He-WD companion. Even for a massive pulsar ($M_{\rm p}$ = $2.0$ M$_\odot$), we have $\overline{M_{\rm c}} = 0.42 M_{\odot}\lesssim 0.5 M_{\odot}$, the rough boundary mass above which instead a CO-WD is expected to form \citep{Tauris1999}. PSR J1216$-$6410 therefore likely has a He-WD companion. It is not possible to make a statement about the likely nature of the companion of PSR J1327$-$0755 since the $M_{\rm c}$ mass ranges associated with a light or heavy pulsar mass easily covers both He-WD and CO-WD solutions, and the mean values obtained for each simulated $M_{\rm p}$ lies close to the divide between He-WD and CO-WD masses.

\begin{table}
\centering
\caption{Constraints (central 68\% confidence interval) on the companion star masses and inclination angles, assuming a pulsar masses of $1.2$ M$\odot$ and $2.0$ M$\odot$. For each system, the value of inclination angle presents an equally probable alternative solution of $180^\circ - \, i$.}
\begin{tabular}{@{}lllll}
\hline
Pulsar & $M_{\rm p}$ ($\rm M_{\odot}$) & $M_{\rm c}$ ($\rm M_{\odot}$) & $\cos i$ & $i^{\circ}$  \\
\hline

J0900$-$3144 & $1.2$ & $0.38$--$0.81$ & $0.48$--$0.87$ & $29$--$62$ \\
             & $2.0$ & $0.55$--$1.13$ & $0.54$--$0.88$ & $28$--$58$ \\

J1216$-$6410 & $1.2$ & $0.19$--$0.37$ & $0.66$--$0.90$ & $25$--$49$ \\  
             & $2.0$ & $0.29$--$0.54$ & $0.72$--$0.91$ & $24$--$44$ \\

J1327$-$0755 & $1.2$ & $0.27$--$0.59$ & $0.63$--$0.91$ & $24$--$51$ \\  
             & $2.0$ & $0.40$--$0.85$ & $0.69$--$0.92$ & $23$--$47$ \\
\hline
\end{tabular}
\label{tab:mass_inclination}
\end{table}

\begin{figure}
\includegraphics[width=0.5\textwidth]{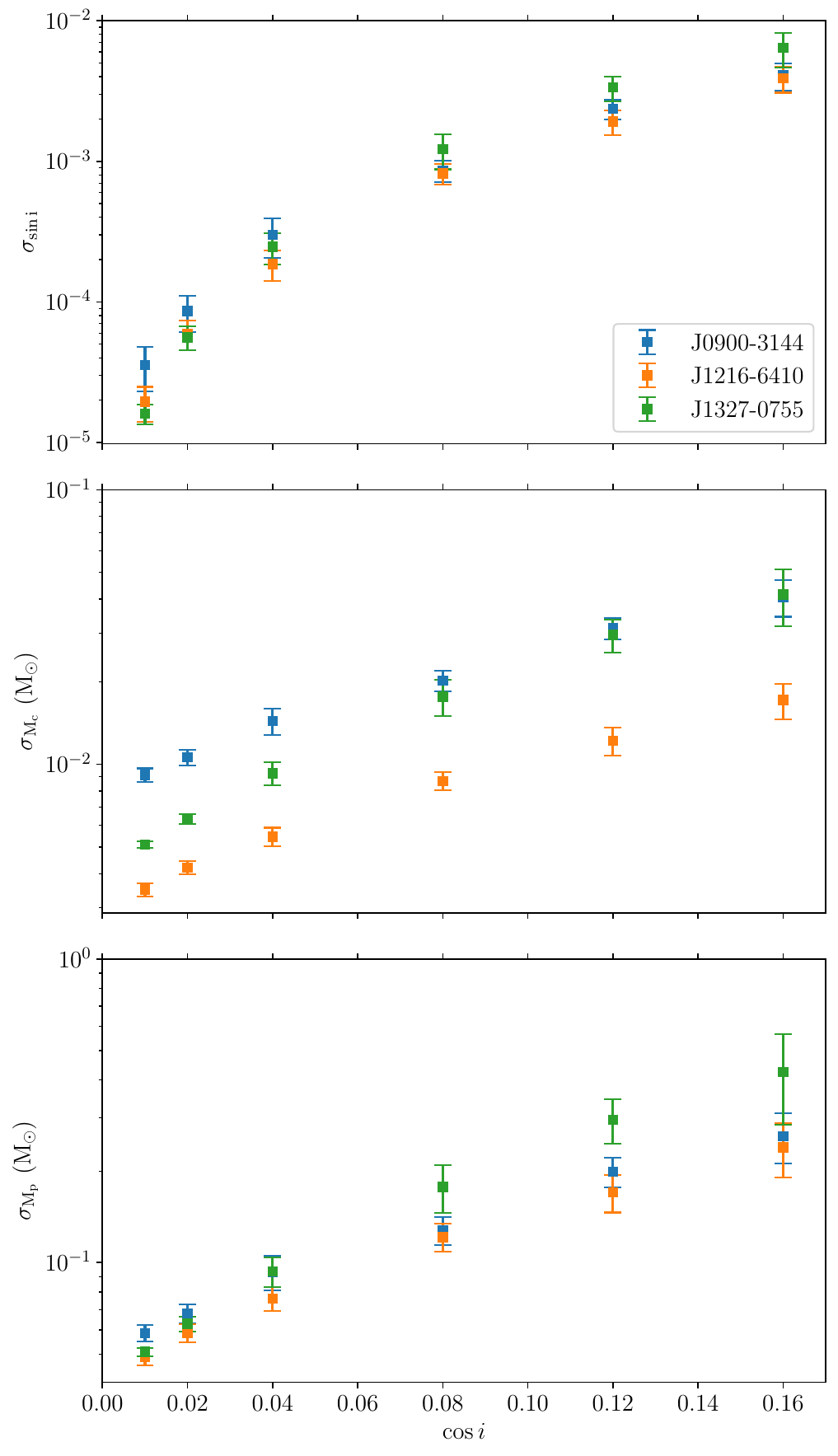}
\caption{Change of uncertainties in $\sin i$ (top panel), $M_{\rm c}$ (middle panel), and $M_{\rm p}$ (bottom panel) as a function of $\cos i$ for PSRs~J0900$-$3144, J1216$-$6410, and J1327$-$0755. }
\label{fig:dsini_dmc_dmp_cosi}
\end{figure}

Figure \ref{fig:dsini_dmc_dmp_cosi} indicates that the these binary MSPs are required to have $\cos i \leq 0.04$ to satisfy the condition of $\sigma_{\rm M_{p}} \leq 0.1 \, {\rm M_{\odot}}$. However, the ranges of $\cos i$ obtained from the $\chi^2$ analysis show that all of these MSPs have $\cos i$ greater than $0.16$. This shows that more observations of these systems will probably not improve the uncertainty of the pulsar mass.

\section{Conclusion}\label{sec:conclusions}

We have updated the timing models for five MSPs using precision timing observations from the MeerKAT radio telescope. We account for noise processes in the timing residuals including pulse jitter, achromatic noise due to spin irregularities in the pulsars or gravitational waves, and dispersion measure variations in the ionised interstellar medium. For PSR~J0900$-$3144 we identified a potential pulsar glitch at the beginning of 2023, but found no significant Bayesian evidence from our data supporting a glitch relative to a model of only red noise. We conclude that further observations are required to determine whether a glitch has occurred. 

For PSR~J0921$-$5202 we presented the first measurement of proper motion by combining the MeerKAT observations with earlier observations from Murriyang to increase the timing baseline. For all pulsars we use the proper motion measurements to derive transverse velocities (Table \ref{tab:timing}), using our derived distances from parallax measurements for PSRs~J1216$-$6410, J1327$-$0755, and J1543$-$5149. We searched for a Shapiro delay in all of the pulsars and report weak detections for PSRs~J0900$-$3144, J1216$-$6410, and J1327$-$0755. We were unable to derive useful constraints on the pulsar mass from these measurements, however, we discuss the likely ranges for the companion masses under assumed pulsar mass values. The companion of PSR~J1216$-$6410 is likely a He-WD and PSR~J0900$-$3144 likely CO-WD. We note that for these systems, future observations are unlikely to produce precise pulsar mass measurements from the Shapiro delay alone.

Finally, we presented the first detection of a clear Shapiro delay in PSR~J1543$-$5149 and use it to derive a precise pulsar mass, $M_{\rm p} = 1.349^{+0.043}_{-0.061}\,$M$_\odot$. The pulsar mass is unusually low for a recycled MSP. The birth mass is estimated to have been $M_{\rm p} \sim 1.25\,$M$_\odot$, making it one of the lightest neutron stars, and the lightest to have formed from the most common MSP formation channel.

\section*{Data Availability}

The pulsar ephemerides, times of arrival, and noise models are available upon request from the corresponding authors. The observations are available from the MeerTime data portal, \url{https://pulsars.org.au/}.

\section*{Acknowledgements}
The MeerKAT telescope is operated by the South African Radio Astronomy Observatory (SARAO), which is a facility of the National Research Foundation, an agency of the Department of Science and Innovation. Funding for the PTUSE machines was provided by the Max-Planck-Institut für Radioastronomie (MPlfR), also supported by the MPG-CAS LEGACY programme, Swinburne University of Technology, and the Australian SKA office. 

This research project was conducted as a student project, following an online MeerKAT pulsar timing workshop (\href{https://www.sarao.ac.za/e-learning-portal/}{https://www.sarao.ac.za/e-learning-portal/}). We thank the organisers and contributors of the workshop and SARAO for its support.  This work was partially undertaken as part of the Australian Research Council Centre of Excellence for Gravitational Wave Discovery (CE170100004 and CE23010016). MG acknowledges support from the South African Research Chairs Initiative of the DSI/NRF (grant No. 98626). Venkatraman Krishnan acknowledges financial support from the European Research Council (ERC) starting grant `COMPACT’ (grant agreement number:101078094). FA acknowledges that part of the research activities described in this paper were carried out with the contribution of the NextGenerationEU funds within the National Recovery and Resilience Plan (PNRR), Mission 4 – Education and Research, Component 2 – From Research to Business (M4C2), Investment Line 3.1 – Strengthening and creation of Research Infrastructures, Project IR0000034 – ‘STILES -Strengthening the Italian Leadership in ELT and SKA’.




\bibliographystyle{mnras}
\bibliography{references} 

\begin{thebibliography}{}
\makeatletter
\relax
\def\mn@urlcharsother{\let\do\@makeother \do\$\do\&\do\#\do\^\do\_\do\%\do\~}
\def\mn@doi{\begingroup\mn@urlcharsother \@ifnextchar [ {\mn@doi@}
  {\mn@doi@[]}}
\def\mn@doi@[#1]#2{\def\@tempa{#1}\ifx\@tempa\@empty \href
  {http://dx.doi.org/#2} {doi:#2}\else \href {http://dx.doi.org/#2} {#1}\fi
  \endgroup}
\def\mn@eprint#1#2{\mn@eprint@#1:#2::\@nil}
\def\mn@eprint@arXiv#1{\href {http://arxiv.org/abs/#1} {{\tt arXiv:#1}}}
\def\mn@eprint@dblp#1{\href {http://dblp.uni-trier.de/rec/bibtex/#1.xml}
  {dblp:#1}}
\def\mn@eprint@#1:#2:#3:#4\@nil{\def\@tempa {#1}\def\@tempb {#2}\def\@tempc
  {#3}\ifx \@tempc \@empty \let \@tempc \@tempb \let \@tempb \@tempa \fi \ifx
  \@tempb \@empty \def\@tempb {arXiv}\fi \@ifundefined
  {mn@eprint@\@tempb}{\@tempb:\@tempc}{\expandafter \expandafter \csname
  mn@eprint@\@tempb\endcsname \expandafter{\@tempc}}}

\bibitem[\protect\citeauthoryear{{Agazie} et~al.,}{{Agazie}
  et~al.}{2023}]{ng23}
{Agazie} G.,  et~al., 2023, \mn@doi [\apjl] {10.3847/2041-8213/acdac6}, \href
  {https://ui.adsabs.harvard.edu/abs/2023ApJ...951L...8A} {951, L8}

\bibitem[\protect\citeauthoryear{{Arzoumanian} et~al.,}{{Arzoumanian}
  et~al.}{2018a}]{Arzoumanian2018}
{Arzoumanian} Z.,  et~al., 2018a, \mn@doi [\apjs] {10.3847/1538-4365/aab5b0},
  \href {https://ui.adsabs.harvard.edu/abs/2018ApJS..235...37A} {235, 37}

\bibitem[\protect\citeauthoryear{{Arzoumanian} et~al.,}{{Arzoumanian}
  et~al.}{2018b}]{Arzoumanian+18}
{Arzoumanian} Z.,  et~al., 2018b, \mn@doi [\apj] {10.3847/1538-4357/aabd3b},
  \href {https://ui.adsabs.harvard.edu/abs/2018ApJ...859...47A} {859, 47}

\bibitem[\protect\citeauthoryear{{Bailes} et~al.,}{{Bailes}
  et~al.}{2020}]{Bailes+20}
{Bailes} M.,  et~al., 2020, \mn@doi [\pasa] {10.1017/pasa.2020.19}, \href
  {https://ui.adsabs.harvard.edu/abs/2020PASA...37...28B} {37, e028}

\bibitem[\protect\citeauthoryear{Barr et~al.,}{Barr et~al.}{2024}]{barr2024}
Barr E.~D.,  et~al., 2024, \mn@doi [Science] {10.1126/science.adg3005}, 383,
  275

\bibitem[\protect\citeauthoryear{Bhattacharya \& van~den Heuvel}{Bhattacharya
  \& van~den Heuvel}{1991}]{bhattacharya1991formation}
Bhattacharya D.,  van~den Heuvel E. P.~J.,  1991, Physics Reports, 203, 1

\bibitem[\protect\citeauthoryear{{Boyles} et~al.,}{{Boyles}
  et~al.}{2013}]{Boyles+13}
{Boyles} J.,  et~al., 2013, \mn@doi [\apj] {10.1088/0004-637X/763/2/80}, \href
  {https://ui.adsabs.harvard.edu/abs/2013ApJ...763...80B} {763, 80}

\bibitem[\protect\citeauthoryear{{Burgay} et~al.,}{{Burgay}
  et~al.}{2006}]{Burgay+06}
{Burgay} M.,  et~al., 2006, \mn@doi [\mnras]
  {10.1111/j.1365-2966.2006.10100.x}, \href
  {https://ui.adsabs.harvard.edu/abs/2006MNRAS.368..283B} {368, 283}

\bibitem[\protect\citeauthoryear{{Choudhury} et~al.,}{{Choudhury}
  et~al.}{2024}]{Choudhury+24}
{Choudhury} D.,  et~al., 2024, \mn@doi [\apjl] {10.3847/2041-8213/ad5a6f},
  \href {https://ui.adsabs.harvard.edu/abs/2024ApJ...971L..20C} {971, L20}

\bibitem[\protect\citeauthoryear{{Cognard} \& {Backer}}{{Cognard} \&
  {Backer}}{2004}]{Cognard+04}
{Cognard} I.,  {Backer} D.~C.,  2004, \mn@doi [\apjl] {10.1086/424692}, \href
  {https://ui.adsabs.harvard.edu/abs/2004ApJ...612L.125C} {612, L125}

\bibitem[\protect\citeauthoryear{{Cordes} \& {Lazio}}{{Cordes} \&
  {Lazio}}{2002}]{ne2001}
{Cordes} J.~M.,  {Lazio} T.~J.~W.,  2002, \mn@doi [arXiv e-prints]
  {10.48550/arXiv.astro-ph/0207156}, \href
  {https://ui.adsabs.harvard.edu/abs/2002astro.ph..7156C} {pp
  astro--ph/0207156}

\bibitem[\protect\citeauthoryear{Cromartie et~al.,}{Cromartie
  et~al.}{2020}]{cromartie2020relativistic}
Cromartie H.~T.,  et~al., 2020, Nature Astronomy, 4, 72

\bibitem[\protect\citeauthoryear{{Desvignes} et~al.,}{{Desvignes}
  et~al.}{2016}]{Desvignes+16}
{Desvignes} G.,  et~al., 2016, \mn@doi [\mnras] {10.1093/mnras/stw483}, \href
  {https://ui.adsabs.harvard.edu/abs/2016MNRAS.458.3341D} {458, 3341}

\bibitem[\protect\citeauthoryear{{EPTA Collaboration} et~al.,}{{EPTA
  Collaboration} et~al.}{2023}]{epta23}
{EPTA Collaboration} et~al., 2023, \mn@doi [\aap]
  {10.1051/0004-6361/202346844}, \href
  {https://ui.adsabs.harvard.edu/abs/2023A&A...678A..50E} {678, A50}

\bibitem[\protect\citeauthoryear{{Edwards}, {Hobbs}  \& {Manchester}}{{Edwards}
  et~al.}{2006}]{Edwards+06}
{Edwards} R.~T.,  {Hobbs} G.~B.,   {Manchester} R.~N.,  2006, \mn@doi [\mnras]
  {10.1111/j.1365-2966.2006.10870.x}, \href
  {https://ui.adsabs.harvard.edu/abs/2006MNRAS.372.1549E} {372, 1549}

\bibitem[\protect\citeauthoryear{Feroz, Hobson  \& Bridges}{Feroz
  et~al.}{2009}]{feroz2009multinest}
Feroz F.,  Hobson M.,   Bridges M.,  2009, Monthly Notices of the Royal
  Astronomical Society, 398, 1601

\bibitem[\protect\citeauthoryear{Fonseca et~al.,}{Fonseca
  et~al.}{2021}]{fonseca2021refined}
Fonseca E.,  et~al., 2021, The Astrophysical Journal Letters, 915, L12

\bibitem[\protect\citeauthoryear{Foster, Backer  \& Wolszczan}{Foster
  et~al.}{1990}]{foster1990timing}
Foster R.,  Backer D.,   Wolszczan A.,  1990, The Astrophysical Journal, 356,
  243

\bibitem[\protect\citeauthoryear{{Freire} \& {Wex}}{{Freire} \&
  {Wex}}{2010}]{FreireAndWex2010}
{Freire} P.~C.~C.,  {Wex} N.,  2010, \mn@doi [\mnras]
  {10.1111/j.1365-2966.2010.17319.x}, \href
  {http://cdsads.u-strasbg.fr/abs/2010MNRAS.409..199F} {409, 199}

\bibitem[\protect\citeauthoryear{{Hotan}, {van Straten}  \&
  {Manchester}}{{Hotan} et~al.}{2004}]{HotanEtAl2004}
{Hotan} A.~W.,  {van Straten} W.,   {Manchester} R.~N.,  2004, \mn@doi [\pasa]
  {10.1071/AS04022}, \href {http://cdsads.u-strasbg.fr/abs/2004PASA...21..302H}
  {21, 302}

\bibitem[\protect\citeauthoryear{Keith et~al.,}{Keith et~al.}{2011}]{Keith2011}
Keith M.~J.,  et~al., 2011, \mn@doi [Monthly Notices of the Royal Astronomical
  Society] {10.1111/j.1365-2966.2011.19842.x}, 419, 1752

\bibitem[\protect\citeauthoryear{{Kramer} et~al.,}{{Kramer}
  et~al.}{2021a}]{KramerGR}
{Kramer} M.,  et~al., 2021a, \mn@doi [Physical Review X]
  {10.1103/PhysRevX.11.041050}, \href
  {https://ui.adsabs.harvard.edu/abs/2021PhRvX..11d1050K} {11, 041050}

\bibitem[\protect\citeauthoryear{{Kramer} et~al.,}{{Kramer}
  et~al.}{2021b}]{Kramer+21}
{Kramer} M.,  et~al., 2021b, \mn@doi [\mnras] {10.1093/mnras/stab375}, \href
  {https://ui.adsabs.harvard.edu/abs/2021MNRAS.504.2094K} {504, 2094}

\bibitem[\protect\citeauthoryear{Lange, Camilo, Wex, Kramer, Backer, Lyne  \&
  Doroshenko}{Lange et~al.}{2001}]{Lange2001}
Lange C.,  Camilo F.,  Wex N.,  Kramer M.,  Backer D.,  Lyne A.,   Doroshenko
  O.,  2001, \mn@doi [Monthly Notices of the Royal Astronomical Society]
  {10.1046/j.1365-8711.2001.04606.x}, 326, 274

\bibitem[\protect\citeauthoryear{Lattimer \& Prakash}{Lattimer \&
  Prakash}{2004}]{lattimer2004physics}
Lattimer J.~M.,  Prakash M.,  2004, science, 304, 536

\bibitem[\protect\citeauthoryear{Lattimer \& Prakash}{Lattimer \&
  Prakash}{2007}]{lattimer2007neutron}
Lattimer J.~M.,  Prakash M.,  2007, Physics reports, 442, 109

\bibitem[\protect\citeauthoryear{{Lentati}, {Alexander}, {Hobson}, {Feroz},
  {van Haasteren}, {Lee}  \& {Shannon}}{{Lentati} et~al.}{2014}]{temponest}
{Lentati} L.,  {Alexander} P.,  {Hobson} M.~P.,  {Feroz} F.,  {van Haasteren}
  R.,  {Lee} K.~J.,   {Shannon} R.~M.,  2014, \mn@doi [\mnras]
  {10.1093/mnras/stt2122}, \href
  {https://ui.adsabs.harvard.edu/abs/2014MNRAS.437.3004L} {437, 3004}

\bibitem[\protect\citeauthoryear{{Lentati} et~al.,}{{Lentati}
  et~al.}{2016}]{Lentati+16}
{Lentati} L.,  et~al., 2016, \mn@doi [\mnras] {10.1093/mnras/stw395}, \href
  {https://ui.adsabs.harvard.edu/abs/2016MNRAS.458.2161L} {458, 2161}

\bibitem[\protect\citeauthoryear{Lorimer et~al.,}{Lorimer
  et~al.}{2006}]{lorimer2006parkes}
Lorimer D.,  et~al., 2006, Monthly Notices of the Royal Astronomical Society,
  372, 777

\bibitem[\protect\citeauthoryear{{Lorimer} et~al.,}{{Lorimer}
  et~al.}{2021a}]{Lorimer+21}
{Lorimer} D.~R.,  et~al., 2021a, \mn@doi [\mnras] {10.1093/mnras/stab2474},
  \href {https://ui.adsabs.harvard.edu/abs/2021MNRAS.507.5303L} {507, 5303}

\bibitem[\protect\citeauthoryear{Lorimer et~al.,}{Lorimer
  et~al.}{2021b}]{lorimer2021timing}
Lorimer D.,  et~al., 2021b, Monthly Notices of the Royal Astronomical Society,
  507, 5303

\bibitem[\protect\citeauthoryear{{Lynch}, {Freire}, {Ransom}  \&
  {Jacoby}}{{Lynch} et~al.}{2012}]{Lynch2012}
{Lynch} R.~S.,  {Freire} P. C.~C.,  {Ransom} S.~M.,   {Jacoby} B.~A.,  2012,
  \mn@doi [\apj] {10.1088/0004-637X/745/2/109}, \href
  {https://ui.adsabs.harvard.edu/abs/2012ApJ...745..109L} {745, 109}

\bibitem[\protect\citeauthoryear{McKee et~al.,}{McKee et~al.}{2016}]{McKee2016}
McKee J.~W.,  et~al., 2016, \mn@doi [Monthly Notices of the Royal Astronomical
  Society] {10.1093/mnras/stw1442}, 461, 2809–2817

\bibitem[\protect\citeauthoryear{Mickaliger et~al.,}{Mickaliger
  et~al.}{2012}]{mickaliger2012discovery}
Mickaliger M.~B.,  et~al., 2012, The Astrophysical Journal, 759, 127

\bibitem[\protect\citeauthoryear{Miles et~al.,}{Miles
  et~al.}{2023}]{miles2023meerkat}
Miles M.~T.,  et~al., 2023, Monthly Notices of the Royal Astronomical Society,
  519, 3976

\bibitem[\protect\citeauthoryear{Miller et~al.,}{Miller
  et~al.}{2021}]{miller2021radius}
Miller M.~C.,  et~al., 2021, The Astrophysical Journal Letters, 918, L28

\bibitem[\protect\citeauthoryear{{NANOGrav Collaboration} et~al.,}{{NANOGrav
  Collaboration} et~al.}{2015}]{ng9yrdata}
{NANOGrav Collaboration} et~al., 2015, \mn@doi [\apj]
  {10.1088/0004-637X/813/1/65}, \href
  {https://ui.adsabs.harvard.edu/abs/2015ApJ...813...65N} {813, 65}

\bibitem[\protect\citeauthoryear{{\"O}zel \& Freire}{{\"O}zel \&
  Freire}{2016}]{ozel2016masses}
{\"O}zel F.,  Freire P.,  2016, Annual Review of Astronomy and Astrophysics,
  54, 401

\bibitem[\protect\citeauthoryear{{Park}, {Folkner}, {Williams}  \&
  {Boggs}}{{Park} et~al.}{2021}]{Park2021}
{Park} R.~S.,  {Folkner} W.~M.,  {Williams} J.~G.,   {Boggs} D.~H.,  2021, \aj,
  161, 105

\bibitem[\protect\citeauthoryear{Parthasarathy et~al.,}{Parthasarathy
  et~al.}{2021}]{parthasarathy2021measurements}
Parthasarathy A.,  et~al., 2021, Monthly Notices of the Royal Astronomical
  Society, 502, 407

\bibitem[\protect\citeauthoryear{Raaijmakers et~al.,}{Raaijmakers
  et~al.}{2021}]{raaijmakers2021constraints}
Raaijmakers G.,  et~al., 2021, The Astrophysical Journal Letters, 918, L29

\bibitem[\protect\citeauthoryear{{Reardon} et~al.,}{{Reardon}
  et~al.}{2023a}]{ppta23}
{Reardon} D.~J.,  et~al., 2023a, \mn@doi [\apjl] {10.3847/2041-8213/acdd02},
  \href {https://ui.adsabs.harvard.edu/abs/2023ApJ...951L...6R} {951, L6}

\bibitem[\protect\citeauthoryear{{Reardon} et~al.,}{{Reardon}
  et~al.}{2023b}]{Reardon+23}
{Reardon} D.~J.,  et~al., 2023b, \mn@doi [\apjl] {10.3847/2041-8213/acdd03},
  \href {https://ui.adsabs.harvard.edu/abs/2023ApJ...951L...7R} {951, L7}

\bibitem[\protect\citeauthoryear{{Reardon} et~al.,}{{Reardon}
  et~al.}{2024}]{Reardon+24}
{Reardon} D.~J.,  et~al., 2024, \mn@doi [\apjl] {10.3847/2041-8213/ad614a},
  \href {https://ui.adsabs.harvard.edu/abs/2024ApJ...971L..18R} {971, L18}

\bibitem[\protect\citeauthoryear{{Ridolfi}, {Freire}, {Gupta}  \&
  {Ransom}}{{Ridolfi} et~al.}{2019}]{Ridolfi2019}
{Ridolfi} A.,  {Freire} P.~C.~C.,  {Gupta} Y.,   {Ransom} S.~M.,  2019, \mn@doi
  [\mnras] {10.1093/mnras/stz2645}, \href
  {https://ui.adsabs.harvard.edu/abs/2019MNRAS.490.3860R} {490, 3860}

\bibitem[\protect\citeauthoryear{Riley et~al.,}{Riley
  et~al.}{2021}]{riley2021nicer}
Riley T.~E.,  et~al., 2021, The Astrophysical Journal Letters, 918, L27

\bibitem[\protect\citeauthoryear{Serylak et~al.,}{Serylak
  et~al.}{2022}]{serylak2022}
Serylak M.,  et~al., 2022, \mn@doi [Astronomy &amp; Astrophysics]
  {10.1051/0004-6361/202142670}, 665, A53

\bibitem[\protect\citeauthoryear{Shamohammadi et~al.,}{Shamohammadi
  et~al.}{2022}]{shamohammadi2022searches}
Shamohammadi M.,  et~al., 2022, Monthly Notices of the Royal Astronomical
  Society

\bibitem[\protect\citeauthoryear{{Shamohammadi} et~al.,}{{Shamohammadi}
  et~al.}{2024}]{shamohammadiastro}
{Shamohammadi} M.,  et~al., 2024, \mn@doi [\mnras] {10.1093/mnras/stae016},
  \href {https://ui.adsabs.harvard.edu/abs/2024MNRAS.530..287S} {530, 287}

\bibitem[\protect\citeauthoryear{Shannon \& Cordes}{Shannon \&
  Cordes}{2010}]{shannon2010assessing}
Shannon R.~M.,  Cordes J.~M.,  2010, The Astrophysical Journal, 725, 1607

\bibitem[\protect\citeauthoryear{{Shapiro}}{{Shapiro}}{1964}]{Shapiro64}
{Shapiro} I.~I.,  1964, \mn@doi [\prl] {10.1103/PhysRevLett.13.789}, \href
  {https://ui.adsabs.harvard.edu/abs/1964PhRvL..13..789S} {13, 789}

\bibitem[\protect\citeauthoryear{{Spiewak} et~al.,}{{Spiewak}
  et~al.}{2022}]{Spiewak+22}
{Spiewak} R.,  et~al., 2022, \mn@doi [\pasa] {10.1017/pasa.2022.19}, \href
  {https://ui.adsabs.harvard.edu/abs/2022PASA...39...27S} {39, e027}

\bibitem[\protect\citeauthoryear{{Stovall} et~al.,}{{Stovall}
  et~al.}{2019}]{Stovall2019}
{Stovall} K.,  et~al., 2019, \mn@doi [\apj] {10.3847/1538-4357/aaf37d}, \href
  {https://ui.adsabs.harvard.edu/abs/2019ApJ...870...74S} {870, 74}

\bibitem[\protect\citeauthoryear{{Tauris} \& {Savonije}}{{Tauris} \&
  {Savonije}}{1999}]{Tauris1999}
{Tauris} T.~M.,  {Savonije} G.~J.,  1999, \mn@doi [\aap]
  {10.48550/arXiv.astro-ph/9909147}, \href
  {https://ui.adsabs.harvard.edu/abs/1999A&A...350..928T} {350, 928}

\bibitem[\protect\citeauthoryear{{Tauris}, {Langer}  \& {Kramer}}{{Tauris}
  et~al.}{2012}]{Tauris+12}
{Tauris} T.~M.,  {Langer} N.,   {Kramer} M.,  2012, \mn@doi [\mnras]
  {10.1111/j.1365-2966.2012.21446.x}, \href
  {https://ui.adsabs.harvard.edu/abs/2012MNRAS.425.1601T} {425, 1601}

\bibitem[\protect\citeauthoryear{{Xu} et~al.,}{{Xu} et~al.}{2023}]{cpta23}
{Xu} H.,  et~al., 2023, \mn@doi [Research in Astronomy and Astrophysics]
  {10.1088/1674-4527/acdfa5}, \href
  {https://ui.adsabs.harvard.edu/abs/2023RAA....23g5024X} {23, 075024}

\bibitem[\protect\citeauthoryear{{Yao}, {Manchester}  \& {Wang}}{{Yao}
  et~al.}{2017}]{YMW17}
{Yao} J.~M.,  {Manchester} R.~N.,   {Wang} N.,  2017, \mn@doi [\apj]
  {10.3847/1538-4357/835/1/29}, \href
  {https://ui.adsabs.harvard.edu/abs/2017ApJ...835...29Y} {835, 29}

\bibitem[\protect\citeauthoryear{{Zhu} et~al.,}{{Zhu} et~al.}{2015}]{Zhu2015}
{Zhu} W.~W.,  et~al., 2015, \mn@doi [\apj] {10.1088/0004-637X/809/1/41}, \href
  {https://ui.adsabs.harvard.edu/abs/2015ApJ...809...41Z} {809, 41}

\makeatother
\end{thebibliography}





\bsp	
\label{lastpage}
\end{document}